\documentstyle[12pt]{article}
\begin{document}\begin{flushright}\thispagestyle{empty}
OUT--4102--85\\
MZ--TH/00--01\\
hep-th/0001202\\
28 January 2000
                \end{flushright}\vspace*{2mm}\begin{center}
                                                   {\large\bf
Towards cohomology of renormalization:\\[4pt]
bigrading the combinatorial Hopf algebra of rooted trees
                                                   }\vglue 10mm{\large\bf
D.~J.~Broadhurst$^{1)}$ and D.~Kreimer$^{2)}$      }\end{center}\vfill
                                                  \noindent{\bf Abstract}\quad
The renormalization of quantum field theory twists the antipode of a
noncocommutative Hopf algebra of rooted trees, decorated by an infinite set of
primitive divergences. The Hopf algebra of undecorated rooted trees,
${\cal H}_R$, generated by a single primitive divergence, solves a universal
problem in Hochschild cohomology. It has two nontrivial closed Hopf
subalgebras: the cocommutative subalgebra ${\cal H}_{\rm ladder}$ of pure
ladder diagrams and the Connes-Moscovici noncocommutative subalgebra
${\cal H}_{\rm CM}$ of noncommutative geometry. These three Hopf algebras
admit a bigrading by $n$, the number of nodes, and an index $k$ that specifies
the degree of primitivity. In each case, we use iterations of the relevant
coproduct to compute the dimensions of subspaces with modest values of $n$ and
$k$ and infer a simple generating procedure for the remainder. The results
for ${\cal H}_{\rm ladder}$ are familiar from the theory of partitions, while
those for ${\cal H}_{\rm CM}$ involve novel transforms of partitions. Most
beautiful is the bigrading of ${\cal H}_R$, the largest of the three. Thanks to
Sloane's {\tt superseeker}, we discovered that it saturates all possible
inequalities. We prove this by using the universal Hochschild-closed
one-cocycle $B_+$, which plugs one set of divergences into another, and by
generalizing the concept of natural growth beyond that entailed by the
Connes-Moscovici case. We emphasize the yet greater challenge of handling the
infinite set of decorations of realistic quantum field theory.
\vfill\footnoterule\noindent
$^1$) D.Broadhurst@open.ac.uk;
http://physics.open.ac.uk/$\;\widetilde{}$dbroadhu\\
Physics Dept, Open University, Milton Keynes MK7 6AA, UK\\
$^2$) Dirk.Kreimer@uni-mainz.de;
http://dipmza.physik.uni-mainz.de/$\;\widetilde{}$kreimer\\
Heisenberg Fellow, Physics Dept, Univ.\ Mainz, 55099 Mainz, Germany
\newpage\setcounter{page}{1}
  \font\tenbb=msbm10
  \newfam\bbfam
  \textfont\bbfam=\tenbb
  \def\bb{\fam\bbfam}
  \def\Qb{{\bb Q}}
  \def\Nb{{\bb N}}
  \newcommand{\dfrac}[2]{\mbox{$\frac{#1}{#2}$}}

\section{Introduction}

In this paper we bigrade the Hopf algebra of undecorated rooted trees,
and both of its closed Hopf subalgebras, taking account of Hochschild
cohomology.

In~\cite{DK}--\cite{shuffle} we have exposed the connection between
renormalization and Hopf algebra. The joblist of renormalization
specifies a noncocommutative coproduct, $\Delta$. On the left are
products of divergent subdiagrams; on the right
these shrink to points. An antipode, $S$, upgrades this bialgebra
to a Hopf algebra, by specifying the procedure of subtracting
subdivergences. If this antipode is twisted, by taking only
the poles of the Laurent series in $\varepsilon$,
for dimensionally regularized diagrams in $d:=4-2\varepsilon$
spacetime dimensions,
the final subtraction delivers a finite renormalized
Green function, in the limit $d\to4$, corresponding
to the minimal subtraction scheme. Different twists
correspond to different renormalization schemes~\cite{DK}.

The general problem of perturbative quantum field theory involves
the Hopf algebra of decorated rooted trees. These decorations
represent primitive divergences, coming from diagrams with no
subdivergences. Restriction to the Hopf algebra ${\cal H}_R$ of
undecorated rooted trees, generated by a single primitive
divergence, reveals a remarkable feature. This apparently small
problem in quantum field theory has a mathematical structure
larger than a very general problem in noncommutative geometry,
investigated by Alain Connes and Henri Moscovici~\cite{CM}, who
showed that the composition of diffeomorphisms can be described
algebraically, and hence extended to noncommutative manifolds, by
making use of an appropriate Hopf algebra ${\cal H}_{\rm CM}$.
In~\cite{CK} it was shown that the Hopf algebra ${\cal H}_{\rm
CM}$ of~\cite{CM} is, in the one-dimensional case, the unique
noncocommutative Hopf subalgebra of ${\cal H}_R$, corresponding to
adding Feynman diagrams~\cite{BK1} with weights determined by
natural growth. The only other closed Hopf subalgebra is the
cocommutative Hopf algebra ${\cal H}_{\rm ladder}$ of rooted trees
whose nodes have fertility less than 2, corresponding to the
ladder (or rainbow) diagrams of~\cite{Hab}--\cite{KD}.

Suppose we are given an $n$-loop Feynman diagram that represents an
$n$-node tree in ${\cal H}_R$. It involves $n$ ultraviolet-divergent
integrations, and hence one may expect that it delivers, in
dimensional regularization, a pole of $n$-th order. But then,
combinations of diagrams corresponding to sums of products
of rooted trees can provide cancellations of poles, and may
hence eliminate leading pole terms. A prominent example of
such a mechanism occurs in the calculation of an anomalous dimension,
$\gamma=d\log Z/d\log \mu$, which detects only single-pole terms,
after minimal subtraction of subdivergences. All higher order poles
are determined by the requirement that they cancel when one takes
the derivative of the logarithm of the renormalization factor $Z$
w.r.t.\ to the renormalization scale $\mu$.
Every practitioner of multiloop quantum chromodynamics
is vividly aware of the bigrading of her/his work,
by loop number and degree of singularity. The slightest error
in handling either the combinatorics or the integrations usually -- and
mercifully -- reveals itself by a failure to get the uniquely
finite answer that is ensured by the locality of counterterms.
Thus there is deep -- and largely uncharted -- structure in the relations
between Laurent expansions of {\em products\/} of Feynman diagrams,
corresponding to {\em forests\/} of rooted trees.

In this work we make preparation
for a cohomological approach to renormalization, by identifying
and analyzing a combinatoric bigrading of
linear combinations of undecorated rooted forests.

In sect.~2, we define this bigrading,
in terms of the number of nodes $n$ and an index
$k$ that classifies subspaces according to their projection into
an augmentation ideal, analyzed by $k$-fold iterations of the
coproduct $\Delta$. We wish to learn the dimension, $H_{n,k}$, of
the subspace with weight $n$ and index $k$. In sect.~3 we find
that this problem has a very simple solution in the cocommutative
subalgebra ${\cal H}_{\rm ladder}$, where the dimension is the
number of ways of partitioning $n$ into $k$ positive integers,
given in Table~1. In sect.~4 we find that the corresponding
problem in the noncocommutative subalgebra ${\cal H}_{\rm CM}$ has
the subtler solution of Table~2, which we find to be related to
Table~1 by a remarkable transform, which preserves the sums of
rows. In sect.~5 we team Neil Sloane's {\tt
superseeker}~\cite{EIS} with Tony Hearn's {\tt Reduce}~\cite{Red}
and find
\begin{equation}
H(x,y):=\sum_{n,k}H_{n,k}x^n y^k={R(x)\over(1-y)R(x)+x y}
\label{ans}
\end{equation}
for the generating function of the bigrading of ${\cal H}_R$,
with results in Table~3 obtained from~\cite{Otter}
\begin{equation}
R(x):=\sum_{n>0}r_n x^n=x\prod_{n>0}(1-x^n)^{-r_n}
=x+x^2+2x^3+4x^4+9x^5+20x^6+\ldots
\label{Rx}
\end{equation}
which generates the number $r_n$ of rooted trees with $n$ nodes.
Our discovery of the generating principle of Table~3
was triggered by {\tt superseeker} analysis of merely the
first 8 entries of its first column. After thorough study of
the filtration in Table~4, we prove~(\ref{ans}).

\section{The second grading}

The weight $n$ of a rooted tree $t$ is the number of its nodes.
The weight of a forest $F=\prod_j t_j$ is the sum of the weights
of the trees $t_j$ in the product. This is the first grading.

To define the index $k$ for the second grading, $k$-primitivity,
we use $k$-fold iterations of the coproduct $\Delta$, defined by the highly
nontrivial recursion~\cite{DK}
\begin{equation}
\Delta(t)=t\otimes e+({\rm id}\otimes B_+)\circ\Delta\circ B_-(t)
\label{D}
\end{equation}
for a nonempty tree $t$. Here $e$ is the empty tree, evaluating to unity,
${\rm id}$ is the identity map, $B_-$ removes the root of $t$, and $B_+$
combines the trees of a product by appending them to a common root.
The coproduct $\Delta$ is coassociative. Hence it has a unique iteration,
which may be written in a variety of equivalent ways.
Since $\Delta$ has only single trees on the right, the recursion
\begin{equation}
\Delta^k=({\rm id}\otimes\Delta^{k-1})\circ\Delta
\label{Dk}
\end{equation}
is particularly convenient. For a forest $F=\prod_j t_j$
we have $\Delta^k(F)=\prod_j\Delta^k(t_j)$.

Let $X$ be a ${\Qb}$-linear combination of monomials of trees,
i.e.~of forests. We
say that $X$ is $k$-primitive if every term of $\Delta^k(X)$ has
at least one empty tree $e$. Symbolically we may consider the
composition of tensor products of the projection operator $P:={\rm
id}-E\circ\bar{e}$ with iterations of $\Delta$. $P$ projects onto
the augmentation ideal ${\cal H}_c=\{X\in{\cal H}_R\mid P(X)=X\}$,
where $X=P(X)+E\circ\bar{e}(X)$. Here
$\bar{e}$ is the counit, which annihilates everything except the
empty tree, for which it gives $\bar{e}(e)=1$. The map from the
rationals back to the algebra is simply $E(q)=q e$ for $q\in{\Qb}$.
Hence $P$ annihilates $e$ and leaves everything else
unchanged. Let $U_0:=P$ and
\begin{equation}
U_k:=(\underbrace{P\otimes\ldots\otimes P}_{k+1~{\rm times}})\circ\Delta^k
=(P\otimes U_{k-1})\circ\Delta
\label{Uk}
\end{equation}
for $k>0$. In using the recursive form, note should be taken
that, in general, the projection makes $U_k(X_1X_2)\not=U_k(X_1)U_k(X_2)$:
one should store results for forests; not just for trees.

We have said that $X$ is $k$-primitive if $U_k(X)=0$. Then clearly $X$ is
$(k+1)$-primitive, since $\Delta^{k+1}(X)$ has at least two
empty trees $e$ in every term. We are interested in the number,
$H_{n,k}:=D_{n,k}-D_{n,k-1}$, of weight-$n$ terms that are
$k$-primitive but are not $(k-1)$-primitive, where $D_{n,k}$
is the dimension of the subspace with weight $n$ and index $k$.
To compute $D_{n,k}$ for specific (and rather modest) values of $n,k$
one considers the most general linear combination $X$ of weight-$n$
terms, with unknown coefficients, and solves $U_k(X)=0$. The rank
deficiency of this large system of linear equations is $D_{n,k}$.
{}From this one subtracts the number $D_{n,k-1}$ of weight-$n$
terms that are $(k-1)$-primitive. By this means we obtained the first
7 rows of Tables~1 and~2, for the Hopf subalgebras ${\cal H}_{\rm ladder}$
and ${\cal H}_{\rm CM}$, and inferred their generating principles.
In the case of the full Hopf algebra ${\cal H}_R$,
bigraded in Table~3, data were much harder to obtain. Fortunately
the generating principle is very distinctive.

\section{Bigrading the cocommutative subalgebra}

We first consider the cocommutative Hopf algebra ${\cal H}_{\rm ladder}$ of
rooted trees all of whose nodes have fertility less than 2,
i.e.~the Hopf algebra with linear basis $l_n=B_+^n(e)$, $n\geq 0$.
In this very simple case, the recursive definition~(\ref{D})
linearizes on the left, giving
\begin{equation}
\Delta(l_n)=\sum_{k=0}^n l_{n-k}\otimes l_k
\label{Dl}
\end{equation}
for the unique $n$-node tree $l_n\in{\cal H}_{\rm ladder}$.
Thanks to our recent work in~\cite{BK2} we have an explicit
construction of the weight-$n$ 1-primitive $p_n\in{\cal H}_{\rm ladder}$.
First we compute the antipodes. In the cocommutative case, these are
simply
\begin{equation}
S(l_n)=-\sum_{k=0}^{n-1}S(l_{n-k})l_k
\label{Sl}
\end{equation}
with $l_0=e$ and $S(e)=e$.
To construct the 1-primitives, we use the star product
$S\star Y$, where $Y$ is the grading operator, giving $Y(l_k)=kl_k$.
In general, a star product of operators is defined by
$O_1\star O_2:=m\circ(O_1\otimes O_2)\circ\Delta$, where $m$ merely
multiplies entries on the left and right of a tensor product.
The ladder 1-primitives are given by
\begin{equation}
p_n:=\frac{1}{n}[S\star Y](l_n)=\sum_{k=0}^{n}\frac{k}{n}
S(l_{n-k})l_k\,.
\label{pl}
\end{equation}
Clearly $p_1=l_1$ and $p_2=l_2-\frac12l_1^2$
are 1-primitive. It takes some time to show that
\begin{eqnarray}
p_8&=&l_8
-l_7l_1
-l_6l_2
+l_6l_1^2
-l_5l_3
+2l_5l_2l_1
-l_5l_1^3
-\dfrac12l_4^2
+2l_4l_3l_1
+l_4l_2^2
-3l_4l_2l_1^2
+l_4l_1^4
\nonumber\\&&{}
+l_3^2l_2
-\dfrac32l_3^2l_1^2
-3l_3l_2^2l_1
+4l_3l_2l_1^3
-l_3l_1^5
-\dfrac14l_2^4
+2l_2^3l_1^2
-\dfrac52l_2^2l_1^4
+l_2l_1^6
-\dfrac18l_1^8
\label{p8}
\end{eqnarray}
gives $\Delta(p_8)=p_8\otimes e+e\otimes p_8$. We were able to compute
this primitive with ease, using recursion~(\ref{Sl})
in the star product~(\ref{pl}).
{}From~\cite{BK2} we know that $[S\star Y](t)$ delivers
a combination of diagrams whose singularity is a single pole, as $d\to4$,
with a residue that determines the contribution of $t$
to the anomalous dimension. Moreover
1-primitives have only single poles. However the converse is not
true in the full Hopf algebra: noncocommutativity implies that
not every $[S\star Y](t)$ is 1-primitive.
Here, in the cocommutative subalgebra,
there is a single 1-primitive for each weight $n>0$.
Hence $S\star Y$ delivers it.

{}From examples such as~(\ref{p8}) we inferred
the general result of~(\ref{pl}).
The 1-primitive $p_n$ contains all
possible multiplicative
partitions $\prod_j l_j^{n_j}$ with weight $n=\sum_j n_j j$.
The coefficient of each partition is $(-1)^{k-1}(k-1)!/\prod_j n_j!$
where $k=\sum_j n_j$ is the number of integers into which
$n$ has been partitioned.
For example the partition $8=2+2+1+1+1+1$, with $k=6$,
gives the coefficient $-5!/2!4!=-5/2$ of $l_2^2l_1^4$ in~(\ref{p8}).
We have tested this Ansatz up to $n=20$, where $p_{20}$ contains
627 terms.

It is easy to understand the leading diagonal of Table~1:
$l_1^n$ is $n$-primitive, but not $(n-1)$-primitive.
For $8>n>k>1$ we used {\tt Reduce} to prove the results of Table~1.
The entry in the $n$-th row and $k$-th column
is $\overline{H}_{n,k}=\overline{D}_{n,k}-\overline{D}_{n,k-1}$,
where $\overline{D}_{n,k}$ is the number of undetermined coefficients
when one solves $U_k(X)=0$, with $X$
taken as an unknown linear combination of forests $\prod_j l_j^{n_j}$
of weight $n=\sum_j n_j j$.
Clearly the generating principle is
extremely simple: $\overline{H}_{n,k}$ is the number of partitions
of $n$ into $k$ positive integers. This simply reflects the fact
that solving $U_k(X)=0$ determines all and only
the coefficients of partitions with $\sum_j n_j\le k$.
Hence the $k$-th column of Table~1 is
generated by
\begin{equation}
\overline{H}_k(x):=
\sum_n\overline{H}_{n,k}x^n=\prod_{j\le k}{x\over1-x^j}
=\frac{x}{1-x^k}\overline{H}_{k-1}(x)
\label{genlad}
\end{equation}
which yields the recursion of the tabular entry {\tt A048789} of~\cite{EIS}:
\begin{equation}
\overline{H}_{n,k}=\overline{H}_{n-k,k}+\overline{H}_{n-1,k-1}
\label{reclad}
\end{equation}
seeded by the empty tree, which gives $\overline{H}_{0,0}=1$.
We particularly note that for all $j,k>0$
\begin{equation}
\overline{H}_{j+k}(x)<\overline{H}_j(x)\overline{H}_k(x).
\label{ltlad}
\end{equation}

\section{Bigrading the Connes-Moscovici subalgebra}

To compute Table~2 we proceeded as above, now
using the coproduct~\cite{CK}
\begin{eqnarray}
\Delta(\delta_n)&=&\delta_n\otimes e+e\otimes\delta_n+R_{n-1}\label{Dd}\\
R_n&=&\left[{\bf X}\otimes e+e\otimes{\bf X}
+\delta_1\otimes{\bf Y},~R_{n-1}\right]+\delta_1\otimes Y(\delta_n)\label{Rd}
\end{eqnarray}
with $R_0=0$, $\left[{\bf X},\delta_n\right]
=\delta_{n+1}$ increasing weight,
and $\left[{\bf Y},\delta_n\right]=Y(\delta_n)=n\delta_n$
measuring weight. This is the noncocommutative coproduct
of Connes and Moscovici~\cite{CM}, shown in~\cite{CK} to give the closed
Hopf subalgebra of ${\cal H}_R$ that is
realized by $\delta_n=N^{n-1}(l_1)$,
where $N$ is the natural growth operator, which appends a single node
in all possible ways.
Thus $\delta_1=l_1$ and $\delta_2=l_2$, while $\delta_3=l_3
+B_+(l_1^2)$ differs from the ladder-algebra element $l_3$.
Natural growth implies that $\delta_n$ is a sum over all weight-$n$
trees in ${\cal H}_R$, with nonzero Connes-Moscovici weights
that we specified in~\cite{BK1}, using an efficient recursive
procedure.

Computation of the first 7 rows of Table~2 took longer than
for Table~1, because of the proliferation of product terms
on the left of the noncocommutative coproduct. These scanty data
presented us with a pretty puzzle,
which the reader might like to try to solve, {\em after\/}
covering up the rows of Table~2 with $n>7$. What is the generating
procedure? Recall that the sum of the $n$-th row in Table~2
must agree with that in Table~1, since each gives the total number of ways
of partitioning the integer $n$. In Table~1 this is achieved with great
simplicity: the $k$-th entry is the number of ways of partitioning $n$
into $k$ positive integers. In Table~2 it is achieved far more subtly,
by the addition of only $1+\lfloor n/2\rfloor$ terms, since
$\widetilde{H}_{n,k}$
has support only for $2k\ge n\ge k$.

Given merely data for $n\le7$,
the most interesting feature is the second subleading diagonal
$1,2,4,6\ldots$ The leading diagonal is
generated by $G_0=1/(1-z)$, the first subleading diagonal by $G_1=1/(1-z)^2$.
The simplest Ansatz for the second is $G_2=G_1/(1-z^2)$,
which requires $\widetilde{H}_{8,6}=9$. Then $\widetilde{H}_{8,5}=4$
is required, so that $1+\widetilde{H}_{8,5}+9+7+1=22$
is the number of ways of partitioning 8.
A {\tt Reduce} program, running
for 24 hours, proved that indeed $\widetilde{H}_{8,5}=4$.
Next, the requirement $\widetilde{H}_{9,6}=7$ comes from $2+
\widetilde{H}_{9,6}+12+8+1=30$, for the partitions of 9,
taking $\widetilde{H}_{9,7}=12$ from
the hypothesis $G_2=1/(1-z)^2(1-z^2)$ for the second subleading diagonal.
Then the third subleading diagonal is revealed as
$1,2,4,7\ldots$ which is nicely consistent with $G_3=G_2/(1-z^3)$.
Finally, it is easy to check that the recurrence relation
$G_k=G_{k-1}/(1-z^k)$ for the diagonals
makes the rows sum to the correct partitions.
Later we shall prove this result by considering the Connes-Moscovici
restriction of the filtration of the bigrading of the full Hopf algebra.

In words, the transformation is simple to state: the subleading diagonals
of Table~2 are the partial sums of the columns of Table~1. This leads to
the subtle recurrence relation
\begin{equation}
\widetilde{H}_{n,k}=\widetilde{H}_{k,2k-n}+\widetilde{H}_{n-2,k-1}
\label{reccm}
\end{equation}
for the bigrading of the Connes-Moscovici Hopf subalgebra.
We particularly note that for $j,k>0$ and $j+k>2$
\begin{equation}
\widetilde{H}_{j+k}(x)<\widetilde{H}_j(x)\widetilde{H}_k(x)
\label{ltcm}
\end{equation}
while for $j=k=1$ we have the equality
$\widetilde{H}_2(x)=\widetilde{H}_1(x)\widetilde{H}_1(x)=x^2(1+x)^2$.

\section{Bigrading the full Hopf algebra of rooted trees}

Given how long it took to compute the
data that eventually led to the generating principle
for the Connes-Moscovici subalgebra, one might be daunted by the task
of inferring the bigrading of the full Hopf algebra of undecorated
rooted trees. In fact, we discovered this first, by mere consideration
of the first 8 entries in the first column of Table~3.
Thanks to~\cite{BK1} we had an extremely
efficient {\tt Reduce} implementation of the coproduct~(\ref{D}).
The severity of the challenge of understanding the range
and kernel of $U_k$, i.e.~the difficulty of the computation
of $\Delta^k$, increases drastically with $k$. At $k=1$ it was
possible to solve $U_1(X):=(P\otimes P)(\Delta(X))=0$, for weights $n\le8$,
using a few hours of CPUtime, notwithstanding the fact that at
$n=8$ the number of products of trees is $r_9=286$.

The book-keeping was very simple, since
the defining property of rooted trees is that every weight-$n$ forest
$F=\prod_j t_j$ is uniquely labelled by the tree $B_+(F)$
with weight $n+1$.
This clearly leads to the enumeration~(\ref{Rx}).
More deeply, it shows that~\cite{CK} $B_+$ is Hochschild closed,
and hence that the apparently simplistic quantum-field-theory task
of handling a single primitive divergence solves a universal problem
in Hochschild cohomology.

Submitting $1,1,1,2,3,8,16,41$ to Neil Sloane's superseeker~\cite{EIS},
we learnt that it is generated by the first 8 terms of
\begin{equation}
H_1(x)=\frac{R(x)-x}{R(x)}=1-\prod_{n>0}(1-x^n)^{r_n}.
\label{H}
\end{equation}
At the time, Sloane had no idea that we were studying
the bigrading of rooted trees and told us
``it is pretty unlikely this is your sequence,
but I thought I should pass this along just in case''.
In fact, his {\tt superseeker} discovery unlocked our puzzle.
We knew that
\begin{equation}
\frac{R(x)}{x}=\sum_{k\ge0} H_k(x)
\label{Rbx}
\end{equation}
where $H_0(x):=1$ and $H_k(x):=\sum_{k}H_{n,k}x^n$ generates column $k$
of Table~3. We then construed~(\ref{H}) as
\begin{equation}
\frac{R(x)}{x}=\frac{1}{1-H_1(x)}=\sum_{k\ge0}[H_1(x)]^k.
\label{const}
\end{equation}
Comparison with~(\ref{Rbx}) then led to the conjecture
$H_k(x)=[H_1(x)]^k$, requiring that
\begin{equation}
H_{j+k}(x)=H_j(x)H_k(x)\,.
\label{req}
\end{equation}
To test this, we made intensive use of {\tt Reduce}.
At weight $n=9$ we computed the $3214\times719$ matrix of integer
contributions to the 3214 terms in $\Delta(X)$ produced by
$r_{10}=719$ weight-9 forests.
The rank deficiency of the condition
$U_1(X)=0$ was proven to be $D_{9,1}=98$, which is indeed the
coefficient of $x^9$ in~(\ref{H}). We tested $H_2(x)=[H_1(x)]^2$
up to weight $n=8$, where $\Delta^2(X)$ has 3651 terms in 286 unknowns.
Here $U_2(X)=0$ gave $D_{8,2}=41+58=99$,
where 41 and 58 are indeed the coefficients of $x^8$ in~(\ref{H})
and its square. Finally, we tested $H_3(x)=H_1(x)H_2(x)$
up to weight $n=7$, where $\Delta^3(X)$ has 3168 terms in 115 unkowns,
with $U_3(X)=0$ giving $D_{7,3}=16+26+27=69$, in agreement
with the sum of the coefficients of $x^7$ in~(\ref{H}), its square
and cube.

Hence we obtained compelling evidence
for the bigrading~(\ref{ans}) of the Hopf algebra
of rooted trees, determined by the circumstance~(\ref{req})
that it {\em saturates all inequalities}. First we derive these general
inequalities, for any commutative graded Hopf algebra.
Then we prove that they are saturated in ${\cal H}_R$.

\subsection{General inequalities}

Let ${\cal H}$ be a commutative graded Hopf algebra
with unit $e$. Let ${\rm deg}$
be the grading, with ${\rm deg}(X)\in {\Nb}$ for all $X\in{\cal H}$
and ${\rm deg}(e)=0$.
We assume that ${\cal H}$ is reduced to scalars by the counit $\bar{e}$.

Let ${\cal H}_k$ be the set of elements in the kernel of $U_k$
which are in the range of $U_{k-1}$, so that $U_k(X)=0$ and
$U_{k-1}(X)\not=0$, for $X\in{\cal H}_k$.
Then we call $k$ the degree of primitivity of $X$, writing
${\rm deg}_p(X)=k$. We let ${\cal H}_0$ be the set of
elements in the kernel of $P=U_0$, i.e.~the scalars. The
augmentation ideal fulfills  ${\cal H}_c={\cal H}/{\cal H}_0=
\sum_{k=1}^\infty {\cal H}_k$.

To show that ${\rm \deg}_p(X)\le{\rm deg}(X)$, suppose that
\begin{equation}
\Delta^{{\rm deg}(X)-1}(X)
=\sum_i X^{(1)}_i\otimes\ldots\otimes X_i^{({\rm deg(X)})}
\label{major}
\end{equation}
has nonscalar entries in ${\cal H}_c^{\otimes {\rm deg}(X)}$.
Then they are all formed from 1-primitives, in ${\cal H}_1$,
since the coproduct is homogenous in ${\rm deg}$
and any element $X$ with ${\rm deg}(X)=1$ also has ${\rm deg}_p(X)=1$.
Hence ${\rm deg}_p$ is majorized by ${\rm deg}$
and is thus finite for each $X\in {\cal H}$.

We denote by $H_{n,k}$ the number of linearly inequivalent terms $X$ with
weight ${\rm deg}(X)=n$ and primitivity ${\rm deg}_p(X)=k$. Then
the generators $H_k(x):=\sum_{n\ge k}H_{n,k}x^n$ satisfy
\begin{equation}
H_{j+k}(x)\le H_j(x)H_k(x),\quad j,k>0\,.
\label{leq}
\end{equation}

{\bf Proof}: It is sufficient to show that an element $X\in{\cal H}_{j+k}$
may be labelled by those terms in $\Delta(X)$ that are in
${\cal H}_j\otimes{\cal H}_k$.
To prove this, suppose that $X_1$ and $X_2$ give the same terms in
${\cal H}_j\otimes{\cal H}_k$. Now observe that $U_{j,k}:=
U_{j-1}\otimes U_{k-1}$ projects onto ${\cal H}_j\otimes{\cal H}_k$,
giving $U_{j,k}\circ\Delta(X_1-X_2)=0$. Finally, observe that
coassociativity gives
\begin{equation}
0=U_{j,k}\Delta(X_1-X_2)=P^{\otimes(j+k)}\circ\Delta^{j+k-1}(X_1-X_2)
:=U_{k+j-1}(X_1-X_2)
\label{done}
\end{equation}
which shows that $X_1-X_2$ is $(j+k-1)$-primitive and hence that
$X_1$ and $X_2$ are equivalent elements of ${\cal H}_{j+k}$.~$\Box$

In consequence of~(\ref{leq}) we obtain
\begin{equation}
H_k(x)\le [H_1(x)]^k\,.
\label{weak}
\end{equation}
This reflects the fact that the terms in $\Delta^{k-1}(X)$
which belong to ${\cal H}_1^{\otimes k}$ are sufficient to label
elements $X\in{\cal H}_k$. The remarkable feature of the
Hopf algebra of rooted trees, to be proved below, is that
all the elements of ${\cal H}_1^{\otimes k}$ are necessary
to label elements of ${\cal H}_k$.

As a further comment, we note that~(\ref{weak}) may be strengthened if
the Hopf algebra is cocommutative, since then the order of labels
is immaterial. In the case of ${\cal H}_{\rm ladder}$, with
$\overline{H}_1(x)=x/(1-x)$, one thus obtains
\begin{equation}
\overline{H}_k(x)\le\prod_{j\le k}{x\over1-x^j}
\label{ladsat}
\end{equation}
which is in fact saturated by Table~1.

\subsection{Saturation}

We now seek to prove that $H_{k}(x)=[H_1(x)]^k$ in the case that
${\cal H}={\cal H}_R$ is the Hopf algebra of undecorated rooted trees.

First we prove that ${\rm deg}_p(X_j X_k)={\rm deg}_p(X_j)
+{\rm deg}_p(X_k)$.

{\bf Proof}: Suppose that $X_j\in{\cal H}_j$
and $X_k\in{\cal H}_k$. Then
\begin{equation}
U_{j+k-1}(X_j X_k)=P^{\otimes(j+k)}
(\Delta^{j+k-1}(X_j)\Delta^{j+k-1}(X_k))\not=0
\label{jk}
\end{equation}
contains 1-primitives in all its slots, giving $U_{j+k}(X_j X_k)=0$,
by coassociativity.~$\Box$

It is instructive to see how this works out for the product $Z X$,
when $Z$ is 1-primitive and $X$ is $k$-primitive. Then
\begin{equation}
\Delta^k(Z)=\sum_{j=1}^{k+1}e\otimes\ldots\otimes
Z_{\mid_{j-\mbox{\tiny th place}}}
\otimes\ldots e
\label{DZ}
\end{equation}
consists of all $k+1$ terms with a single $Z$ and $k$ empty trees.
As $X$ is $k$-primitive,
\begin{equation}
\Delta^{k}(X)=\sum_{j=1}^{k+1}\sum_{i_j}
X_{i_j}^{(1)}\otimes\ldots\otimes e_{\mid_{j-\mbox{\tiny th place}}}
\otimes\ldots X_{i_j}^{(k+1)}+\ldots
\label{DX}
\end{equation}
with the final ellipsis denoting omission of terms that contain more than
one $e$. The latter make no contribution to
\begin{equation}
U_k(Z X)
=\sum_{j=1}^{k+1}\sum_{i_j} X_{i_j}^{(1)}\otimes\ldots\otimes
X_{i_j}^{(j-1)}\otimes Z_{\mid_{j-\mbox{\tiny th place}}}\otimes
X_{i_j}^{(j+1)}\otimes\ldots\otimes X_{i_j}^{(k+1)}
\label{UZX}
\end{equation}
where $Z$ replaces a single $e$.
By construction~(\ref{UZX})
has all its entries, namely $Z$ or $X_{i_j}^{(r)}$, in ${\cal H}_1$.
Hence $U_{k+1}(Z X)=0$ and ${\rm deg}_p(Z X)=k+1$.

Iterating this result one immediately
concludes that ${\rm deg}_p(X_1\ldots X_k)=k$,
for 1-primitive elements $X_1,\ldots,X_k$.
This does not, of itself, allow us to conclude that
$H_{k}(x)=[H_1(x)]^k$, since the products are commutative.
Thus there are fewer $k$-fold products of 1-primitives than there
are $k$-primitives.

To appreciate what is needed in the next step,
we pause to consider ${\cal H}_{\rm CM}$. Its 1-primitives are
$\delta_1$ and $\widetilde\delta_2=\delta_2-\frac{1}{2}\delta_1^2$.
{}From these we can form the 2-primitive products
$\delta_1^2$, $\delta_1\widetilde\delta_2$,
and $\widetilde\delta_2^2$.
Table~2 shows that there is a further inequivalent 2-primitive,
at weight $n=3$. Direct computation shows that it may be taken
as $\widetilde\delta_3=\delta_3-\frac{1}{2}\delta_1^3$. Then we may form
6 inequivalent 3-primitive products, namely $\delta_1^3$,
$\delta_1^2\widetilde\delta_2$, $\delta_1\widetilde\delta_3$,
$\delta_1\widetilde\delta_2^2$, $\widetilde\delta_2\widetilde\delta_3$
and $\widetilde\delta_2^3$. Table~2 shows that there is only
one more 3-primitive, at weight $n=4$. It may be taken as
$\widetilde\delta_4=\delta_4-\frac34\delta_1^4$.
The absence of a further inequivalent 3-primitive
at weight $n=5$ means that
$\widetilde{H}_3(x)<\widetilde{H}_1(x)\widetilde{H}_2(x)$.
This exercise
reveals the filtration of the bigrading of Table~2:
the generator is
\begin{equation}
\sum_{n,k}\widetilde{H}_{n,k}x^n y^k={1\over1-x y}
\prod_{k>0}{1\over1-x^{k+1}y^k}
\label{ficm}
\end{equation}
corresponding to products of $l_1$ and
\begin{equation}
\widetilde\delta_{k+1}=N^k(l_1)-\frac{k!}{2^k}l_1^{k+1}
\label{sub}
\end{equation}
with $k$-primitivity achieved by a subtraction at weight $n=k+1>1$.
At $y=1$, the filtration~(\ref{ficm}) agrees with the
ladder filtration
\begin{equation}
\sum_{n,k}\overline{H}_{n,k}x^n y^k=\prod_{k>0}{1\over1-x^k y}
\label{filad}
\end{equation}
generated by products of the 1-primitives~(\ref{pl}).

Now consider the highly nontrivial filtration of the
bigrading of rooted trees.
Let $P_{n,k}$ be the number of weight-$n$ elements of ${\cal H}_k$
that cannot be expressed as products of
elements of $\{{\cal H}_j\mid j<k\}$. Then
\begin{equation}
H(x,y):=\sum_{n,k}H_{n,k} x^n y^k=\prod_{n,k}{1\over(1-x^n y^k)^{P_{n,k}}}
\label{filter}
\end{equation}
with $P_{n,k}$ telling us how many linearly independent combinations of
weight-$n$ trees may be made $k$-primitive, but not $(k-1)$-primitive,
by suitable subtractions of products of trees of lesser weight.
Setting $y=1$, taking logs, and using the unique property~(\ref{Rx})
of the enumeration of rooted trees, we obtain
\begin{equation}
\sum_n r_n \log(1-x^n)=\log x-\log R(x)
=\sum_{n,k} P_{n,k} \log(1-x^n)
\label{assume}
\end{equation}
and hence $r_n=\sum_k P_{n,k}$.

Table~4 gives the filtration implied by $H_k(x)=[H_1(x)]^k$. The
column generators are
\begin{equation}
P_k(x):=\sum_n P_{n,k}x^n
=\sum_{j|k}{\mu(j)\over k}\left(1-\prod_n(1-x^{nj})^{r_n}\right)^{k/j}
\label{mu}
\end{equation}
where the M\"obius function $\mu(j)$ vanishes if $j$ is divisible
by a square and is equal to $(-1)^p$
when $j$ is the product of $p$ distinct primes.

To proceed, we use the Hochschild property of $B_+$, namely
\begin{equation}
\Delta\circ B_+=B_+\otimes e+({\rm id}\otimes B_+)\circ\Delta
\label{BP}
\end{equation}
which follows from the action of the coproduct~(\ref{D}) on the
trees produced by $B_+$, using $B_-\circ B_+={\rm id}$.
Taking care to note that
\begin{equation}
C:=B_+\circ B_-\not=B_-\circ B_+={\rm id}
\label{C}
\end{equation}
we obtain
\begin{equation}
\Delta\circ B_-=({\rm id}\otimes B_-)\circ\Delta\circ C
\label{BM}
\end{equation}
by composition of~(\ref{BP}) with ${\rm id}\otimes B_-$ on the left
and $B_-$ on the right.
It follows from~(\ref{BP}) that if $X$ is $k$-primitive,
then $B_+(X)$ has primitivity no greater than $k+1$.

{\bf Proof}: Suppose that $X\in{\cal H}_k$. Then repeated application
of~(\ref{BP}) gives
\begin{equation}
U_{k+1}\circ B_+(X)
=(P^{\otimes(k+1)}\otimes B_+)\circ\Delta^{k+1}(X)=0
\label{UBP}
\end{equation}
since every term in $\Delta^{k+1}(X)$ contains at least two $e$'s,
of which at most one is promoted to $l_1$ by $B_+$.~$\Box$

The presence of $C$ in~(\ref{BM}) frustrates
a parallel attempt to show that $B_-$ decreases primitivity.
Rather, we found that the kernel of $B_-$ is an object of great interest.
The action of $B_-$ on a nonempty tree $t$ is simple:
it removes the root to produce,
in general, a forest of rooted trees, each of whose roots was originally
connected to the root of $t$ by a single edge. Since $B_-$
obeys the Leibniz rule
\begin{equation}
B_-(X_1 X_2)=X_1 B_-(X_2)+X_2 B_-(X_1),\quad B_-(e)=0,
\label{Lz}
\end{equation}
its action on forests is less trivial. The action of $B_-$ on a tree, $t$,
is undone by $B_+$, giving $C(t):=B_+(B_-(t))=t$. On a forest of more
than one tree, $C$ does not degenerate to the identity map. It is this
that makes the Hopf algebra of rooted trees such an amazingly rich
structure. Another important feature is that the kernels of
$B_-$ and $C$ coincide, since $C:=B_+\circ B_-$ and $B_-=B_-\circ C$.
Moreover, $C$ is idempotent, since
\begin{equation}
(C-{\rm id})\circ C=
B_+\circ(B_-\circ B_+-{\rm id})\circ B_-=0.
\label{idem}
\end{equation}
Hence there
are two special types of object: trees, for which $C$ acts like
the identity, and those linear combinations of forests that lie in the
kernel of $C$. We shall show that the latter are the key to the
filtration $P_{n,k}$ of Table~4. The first step is to prove
that $C(X)=0$ for every $X\in{\cal H}_1$ with weight $n>1$.

{\bf Proof}: The coproduct of tree $t$ has the form
\begin{equation}
\Delta(t)=t\otimes e+B_-(t)\otimes l_1+\ldots
\label{l1t}
\end{equation}
where the ellipsis denotes terms with weight $n>1$ on the right.
Now consider a forest $F=\prod_j t_j$. The Leibniz rule~(\ref{Lz})
gives
\begin{equation}
\Delta(F)=\prod_j \Delta(t_j)=F\otimes e+B_-(F)\otimes l_1+\ldots
\label{l1F}
\end{equation}
and hence $\Delta(X)$ contains $B_-(X)\otimes l_1$, for all $X\in{\cal H}_R$.
Now suppose that $X\in{\cal H}_1$ has no weight-1 term.
Then $\Delta(X)=X\otimes e+e\otimes X$ requires that $B_-(X)=0$
and hence that $X$ is in the kernel of $C$.~$\Box$

To get acquainted with the problem in hand,
consider a pair of 1-primitives, $X_1$ and $X_2$.
Their product is 2-primitive,
giving
\begin{equation}
U_1(X_1 X_2)=X_1\otimes X_2+X_2\otimes X_1\,.
\label{U1s}
\end{equation}
For every such pair, we require another 2-primitive construct,
say $W(X_1,X_2)$,
giving
\begin{equation}
U_1\circ W(X_1,X_2)=X_1\otimes X_2-X_2\otimes X_1\,.
\label{U1a}
\end{equation}
This does not uniquely define $W(X_1,X_2)$, since we may add to any solution
of~(\ref{U1a}) any combination of 1-primitives. The operative question
is whether a solution exists, for each pair of distinct 1-primitives.
This question does not arise in the ladder subalgebra, which is
cocommutative. It is easily answered in the Connes-Moscovici
subalgebra, where the asymmetry of
\begin{equation}
U_1(\widetilde\delta_3)=3\delta_1\otimes\widetilde\delta_2+
\widetilde\delta_2\otimes\delta_1
\label{CMasy}
\end{equation}
makes it simple to solve the single case of~(\ref{U1a}) by
\begin{equation}
W(\delta_1,\widetilde\delta_2)=\widetilde\delta_3
-2\widetilde\delta_2\delta_1
=B_+(l_1^2)+l_3-2l_2l_1+\dfrac12l_1^3.
\label{CMsol}
\end{equation}
More generally, the $k$-primitive nonproduct term
$\widetilde\delta_{k+1}$
accounts for the leading diagonal $P_{k+1,k}=1$ of Table~4.

In the full Hopf algebra, we must show the existence of $P_{n,2}$
asymmetric pairings enumerated by
\begin{equation}
P_2(x):=\sum_n P_{n,2} x^n
=\dfrac12[H_1(x)]^2-\dfrac12H_1(x^2)
=x^3+x^4+3x^5+5x^6+13x^7+28x^8+\ldots
\label{P2}
\end{equation}
Part of what is required is clearly provided by
\begin{equation}
W(l_1,X)=l_1 X-2B_+(X)
\label{WX1}
\end{equation}
since~(\ref{BP}) shows that
\begin{equation}
U_1\circ W(l_1,X)=l_1\otimes X+X\otimes l_1
-2(P\circ B_+)(X\otimes e+e\otimes X)=l_1\otimes X-X\otimes l_1
\label{U1X}
\end{equation}
has the desired antisymmetry. By this means we easily construct the
elements of ${\cal H}_2$ with weight $n<5$ from products
of 1-primitives and the action of $B_+$ on 1-primitives.

At weights $n\ge5$ we need a further construction.
There are $P_{5,2}=3$ weight-5 nonproduct 2-primitives,
but only $H_{4,1}=2$ weight-4 1-primitives on which to act with $B_+$.
We lack, thus far, a way of constructing $W(p_2,p_3)$,
where $p_2=l_2-\frac12l_1^2$ and $p_3=l_3-l_2l_1+\frac13l_1^3$
are the 1-primitives at weights $n=2,3$, common
to the cocommutative subalgebra ${\cal H}_{\rm ladder}$.
At weight $n=6$, we lack $W(p_2,p_4)$ and $W(p_2,p^\prime_4)$,
where
\begin{eqnarray}
p_4&=&l_4-l_3l_1-\dfrac12l_2^2+l_2l_1^2-\dfrac14l_1^4,\label{p4}\\
p^\prime_4&=&B_+\left(2l_2l_1-B_+(l_1^2)-l_1^3\right)
+l_1B_+(l_1^2)-l_2^2,\label{q4}
\end{eqnarray}
are the weight-4 ladder and nonladder 1-primitives enumerated
by $H_{4,1}=2$. It is simple to check that they are annihilated by
the Leibniz action of $B_-$, using $B_-\circ B_+={\rm id}$ and
$B_-(l_n)=l_{n-1}$ with $l_0:=e$ evaluating to unity.

At this juncture, it is instructive to compare
Tables~3 and~4, which reveal that
\begin{eqnarray}
P_{n,2}&\le&2H_{n-1,1}\,\label{keq2}\\
P_{n,k}&\le&H_{n-1,k-1}\,,\quad k>2\label{kgt2}.
\end{eqnarray}
In the Appendix, we show that these inequalities persist at large $n$,
thanks to the fact that the Otter constant
$c:=\lim_{n\to\infty} r_{n+1}/r_n=2.955765\ldots$
is slightly less than 3.
Thus it is conceivable that for $k>2$ the action of $B_+$ might generate
$P_k(x)$ from $xH_{k-1}(x)$, but it is quite impossible
for it to do this job at $k=2$. It appears from~(\ref{keq2})
that we need a second operator that increases $n$ and $k$ by unity.

\subsection{Natural growth by a single node}

There is a clear candidate for the second operator:
the natural growth operator $N$, which appends a single node
in all possible ways, and hence obeys a Leibniz rule.

The commutators of $N$ with $B_\pm$ are easily found, since
we need only consider what is happening at the root. Defining
the operator $L$ by $L(X):=l_1X$, we obtain
\begin{eqnarray}
\left[N,\,B_+\right]&=&B_+\circ L\,,\label{np}\\
\left[B_-,\,N\right]&=&L\circ B_-\,,\label{nm}\\
\left[N,\,C\right]&=&[N,B_+]\circ B_--B_+\circ[B_-,N]=0\,.\label{nc}
\end{eqnarray}
The natural growth operator is a wonderful
thing: it commutes with $B_+\circ B_-$,
the operator that makes the Hopf algebra so structured;
hence it preserves the kernel of $B_-$;
like $B_-$, it acts as a derivative;
like $B_+$, it adds a node and increases the degree of primitivity;
finally, it identifies the unique~\cite{CK}
noncocommutative Hopf subalgebra ${\cal H}_{\rm CM}$,
with linear basis $\delta_{n}:=N^{n-1}(l_1)$.

Constructing $N(p_4)$ and $N(p_4^\prime)$, we verified that they
are in the kernel of $U_2$ and the range of $U_1$. It might thus
appear that some linear combination of them with $B_+(p_4)$,
$B_+(p_4^\prime)$ and the product terms $\{p_1p_4,p_1p_4^\prime,
p_2p_3\}$ solves the problem of constructing $W(p_2,p_3)$.
Remarkably, this turns out not to be the case. Rather, we find that
application of
\begin{equation}
S_1:=N+(B_+-L)\circ Y
\label{S1}
\end{equation}
to a 1-primitive gives a 1-primitive of higher weight.
Here $Y$ is the grading operator, which multiplies each tree by its weight
and operates on products by a Leibniz rule.
Thus $N(p_4)$ and $N(p_4^\prime)$ are linear combinations
of $\{B_+(p_4),B_+(p_4^\prime),p_1p_4,p_1p_4^\prime\}$
and 1-primitives. Instead of constructing
the missing weight-5 nonproduct 2-primitive, we discovered how to
generate all the 1-primitives with weight $n\le5$. We have
$L(e)=p_1=l_1$, at $n=1$; $S_1(p_1)=2p_2$, at $n=2$;
$S_1(p_2)=3p_3$, at $n=3$. At $n=4$, we obtain $S_1(p_3)=4p_4-p_4^\prime$,
to which we adjoin $p_4$, from the ladder construction~(\ref{pl}) of sect.~3.
Then we obtain the 1-primitives at $n=5$ as $p_5$, $S_1(p_4)$ and
$S_1^2(p_3)$.

We then found a generalization of~(\ref{S1}), which solves
the problem of constructing $W(p_2,p_3)$.
Operating on a weight-$n$ 1-primitive with
\begin{equation}
S_k:=\left(S_1-\frac{k-1}{2}(B_+-L)\right)\circ N^{k-1}
\label{Sk}
\end{equation}
we create a $k$-primitive of weight $n+k$.
In particular,
\begin{equation}
W(p_2,p_3)=\frac{8S_2(p_3)-7N\circ S_1(p_3)}{12}-p_2p_3
\label{W23}
\end{equation}
completes the construction of weight-5 2-primitives.
More generally, we found that
\begin{eqnarray}
W(p_2,X_n)&=&\frac{2O_2(X_n)}{n(n+1)}-p_2X_n\label{W2n}\\
O_2&:=&S_2\circ(Y+{\rm id})-N\circ S_1\circ(Y+\dfrac12{\rm id})\label{O2}
\end{eqnarray}
gives $U_1\circ W(p_2,X_n)=p_2\otimes X_n-X_n\otimes p_2$,
where $X_n$ is a 1-primitive with weight $n$. We remark
that~(\ref{W2n}) lies in the kernel of $C$, for all $n>1$.

However, it is not yet clear how to generalize this construction
to obtain, for example, $W(p_3,p_4)$ and $W(p_3,p_4^\prime)$ at weight $n=7$.
They key to this issue is an extension\footnote{Our extension
of natural growth allows a suitable
extension of the Lie algebra dual to ${\cal H}_R$, as was observed
by Alain Connes. This will be presented in a sequel to~\cite{DR2}.}
of the concept of natural growth.

\subsection{Natural growth by appending sums of forests}

Let $F$ be a forest. We define $N_F(X)$ to be the sum of forests
obtained by appending $F$ to every node of $X$, in turn. To append
$F=\prod_j t_j$ to a particular node, one connects the roots of
all the $t_j$ to that node. We note that $N_F$ obeys a Leibniz
rule, with $N_F(e)=0$ and $N_F(l_1)=B_+(F)$. We have already
encountered two examples, namely the grading operator $Y:=N_{e}$,
which merely counts nodes, and the simplest natural growth
operator $N:=N_{l_1}$, which appends a single node. Finally, with
$Z=F_1+F_2$, we make $N_Z:=N_{F_1}+N_{F_2}$ linear in its
subscript, as well as its argument.

The commutation relations~(\ref{np},\ref{nm}) then generalize to
\begin{eqnarray}
\left[N_Z,\,B_+\right]&=&B_+\circ L_Z\label{npx}\\
\left[B_-,\,N_Z\right]&=&L_Z\circ B_-\label{nmx}
\end{eqnarray}
with $L_Z(X):=Z X$. Thus
$[N_Z,C]=0$ and $N_Z$ preserves the kernel of $C$
for all $Z\in{\cal H}_R$.

The great virtue of this construct is that it gives
\begin{equation}
U_1\circ N_{Z}(X)=Z\otimes Y(X)
\label{H2}
\end{equation}
when both $Z$ and $X$ are 1-primitive.

{\bf Proof}: We use the shorthand notation
$\Delta(X)=X\otimes e +e\otimes X+X^\prime\otimes X^{\prime\prime}$
for any Hopf algebra element $X$, with the final
term denoting a sum over tensor products containing no scalars.
Let $Z$ be any 1-primitive. Then
\begin{equation}
U_1\circ N_Z(X)=
N_Z(X^\prime)\otimes X^{\prime\prime}
+X^\prime\otimes N_Z(X^{\prime\prime})
+(L_Z\otimes Y)\circ\Delta(X)
\label{UNZ}
\end{equation}
consists of terms in which $X^\prime$ or $X^{\prime\prime}$
grow naturally, with a final contribution
where $Z$ is itself completely cut from any node
to which it was connected by $N_Z$, with the grading operator $Y$
acting on the right, to count the number of cuts.
The case with $Z=l_1$ was proven in~\cite{CK},
by an analysis of admissible cuts.
Here, where $Z$ is 1-primitive, we obtain a result of the same form,
since the internal cuts of $Z$ cancel when $U_1(Z)=0$.
(A more general formula, for arbitrary $Z$, can be given
but is not required here.)
When $X$ is 1-primitive, with $X^\prime=X^{\prime\prime}=0$,
we obtain~(\ref{H2}) from $L_Z\otimes Y$ acting on the second term of
$\Delta(X)=X\otimes e+e\otimes X$.~$\Box$

The result~(\ref{H2}) immediately proves that $H_2(x)=[H_1(x)]^2$,
since it shows that each pairing $N_{X_1}(X_2)$ of 1-primitives
gives an element of ${\cal H}_2$ that is inequivalent to any other
pairing. Hence~(\ref{weak}) is saturated at $k=2$. Now we define
the iteration
\begin{equation}
V_{k+1}(X_1,\ldots,X_k,X_{k+1}):=N_{V_k(X_1,\ldots,X_k)}(X_{k+1})
\label{Vk}
\end{equation}
for $k>0$, with $V_1:={\rm id}$.
Then, for example, $V_2(X_1,X_2):=N_{X_1}(X_2)$ and
\begin{equation}
V_3(X_1,X_2,X_3):=N_{N_{X_1}(X_2)}(X_3)
\not=N_{X_1}\left(N_{X_2}(X_3)\right)\,.
\label{V3}
\end{equation}
We remark that a Hochschild boundary can be defined for maps
$V_{k+1}:{\cal H}_R^{\otimes(k+1)}\to{\cal H}_R$. For this,
it is sufficient to define terms of the form
$V_k(X_1,\ldots,X_j X_{j+1},\ldots,X_{k+1})$,
where one argument is a product. Natural growth by forests supplies this.
Consequences will be described in future work.
For the present, we are content with the following result.

{\bf Theorem}: The dimensions $H_{n,k}$
of the bigrading of the Hopf algebra of undecorated
rooted trees, by weight $n$ and degree of primitivity $k$,
are generated by~(\ref{ans}).

{\bf Proof}: Let $X_1,X_2,\ldots,X_k$ be 1-primitives, which need not
be distinct. Then
\begin{equation}
U_{k-1}\circ V_k(X_1,X_2,\ldots,X_k)=
X_1\otimes Y(X_2)\otimes\ldots\otimes Y(X_k)
\label{Ukm}
\end{equation}
by coassociativity and iteration of the argument that led to~(\ref{H2}).
Thus $H_k(x)=[H_1(x)]^k$ saturates~(\ref{weak}). Then $H(x,y)=1/(1-H_1(x)y)$
gives $R(x)=x/(1-H_1(x))$, at $y=1$.
Solving for $H_1(x)=1-x/R(x)$, we obtain~(\ref{ans}).~$\Box$

\subsection{Comments on the main theorem}

Four comments are in order. The first concerns the enumeration of
the filtration. This follows from taking logs in~(\ref{filter}),
which gives
\begin{equation}
\log H(x,y)=-\log(1-H_1(x)y)=-\sum_{n,k} P_{n,k}\log(1-x^n y^k)\,.
\label{logs}
\end{equation}
Equating coefficients of $y^j$, and setting $x=z^{1/j}$, we obtain
\begin{equation}
[H_1(z^{1/j})]^j=\sum_{k|j} k P_k(z^{1/k})
\label{classic}
\end{equation}
which is a classic problem in M\"obius inversion, yielding~(\ref{mu}),
after use of~(\ref{Rx}).

Next, we remark on the number, $C_{n,k}$,
of weight-$n$ elements of ${\cal H}_k$ that are in the kernel of
$C:=B_+\circ B_-$. We have explicitly constructed
a filtration of the bigrading, for weights $n<7$,
in which the only element with $C(X)\not=0$ is $l_1$.
The iteration~(\ref{Vk}) proves that there is no obstacle
to continuing this process, since the only restriction
imposed by $C\circ V_{k+1}(X_1,X_2,\ldots,X_{k+1})=0$
is $X_{k+1}\not=l_1$.
Thus $\sum_n C_{n,k+1}x^n=[H_1(x)]^k(H_1(x)-x)$
and the generating function
\begin{equation}
\sum_{n,k}C_{n,k}x^n y^k={(1-x y)R(x)\over(1-y)R(x)+x y}
\label{cans}
\end{equation}
differs from~(\ref{ans}) only by a factor of $1-x y$, which
removes $l_1$ from the filtration~(\ref{filter}).
In total, we have $C_n:=\sum_k C_{n,k}=r_{n+1}-r_n$ weight-$n$ solutions
to $C(X)=0$. It is easy to see how that comes about:
there are $r_{n+1}$ possible forests in $X$, subject to the $r_n$
conditions that the coefficient of every tree in
$C(X)$ vanishes. The result $C_n=r_{n+1}-r_n$ proves
the independence of these conditions.
Hence an element $X$ of the kernel of $C$ is uniquely identified
by the contribution $\overline{X}$ that contains no pure trees,
since $X=\overline{X}-C(\overline{X})$.
Finally, the filtration of the bigrading of the kernel of $C$
differs from that of the full Hopf algebra only by the absence of $l_1$.
These distinctive features frustrate every attempt
to decrease primitivity by the action of $B_-$ on any
nonproduct element except the single-node tree.
One may climb up the ladder of primitivity with great ease,
yet descent is impossible, save in one trivial case.
In a sense, the second grading is characterized
by the profound difficulty of constructing its 1-primitives.
At first meeting, this makes it difficult to fathom. Then
one realizes that the structure is beautifully tuned to prevent
casual construction.

Our third comment concerns the remarkable
operator $O_2$ in~(\ref{O2}),
which provides a way of solving $U_1\circ W(p_2,X)=
p_2\otimes X-X\otimes p_2$. A second way is provided by $N_{p_2}$.
These solutions need not be the same;
they may differ by a 1-primitive. In general, they will differ,
since $N_{p_2}$ acts by a Leibniz rule, while $O_2$ does not.
Hence
\begin{equation}
T_2:=O_2-N_{p_2}\circ(Y+{\rm id})
\label{T1}
\end{equation}
provides a second shift operator that creates 1-primitives,
when applied to 1-primitives.
It gives information that is not provided
by $S_1$ in~(\ref{S1}). For example,
at weight $n=6$ we already know how to construct 4 of the $H_{6,1}=8$
primitives, by applying powers of $S_1$ to the ladder primitives
constructed in~(\ref{pl}). Of the missing 4, the constructs
$T_2(p_4)$ and $T_2\circ S_1(p_3)$ provide 2.
For the remaining 2, which are now proven to exist,
we laboriously solved $U_1(X)=0$ at weight $n=6$,
working with tensor products of the 38 forests with up to 6 nodes.
At first sight,
one might hope to add a few more shift operators,
to arrive at a set that is sufficient to construct
1-primitives up to some large weight, without
having to solve the fearsome explosion of linear equations
required by the vanishing of all tensor products
in $U_1(X)=0$.  This seems not to be the case; the construction
of 1-primitives appears to be a deeply nontrivial challenge.
Asymptotically, no more than a fraction $1/c$ of what is necessary
may be provided by $S_1$, and no more than $1/c^2$ by $T_2$,
which increases weight by 2 units. The number of similarly
constructed operators that change weight by $n$ cannot exceed
the number $H_{n,1}$ of weight-$n$ 1-primitives.
Constructing a finite number of these, we obtain merely an asymptotic
fraction $f<H_1(1/c)=1-1/c<1$ of what is needed.
Hence we envisage no easy route to the construction of 1-primitives,
short of solving the
tensorial defining property $\Delta(X)=X\otimes e+e\otimes X$.
Thereafter, the problem of constructing $k$-primitives is completely
solved by~(\ref{Vk}), which shows that the 1-primitives of weight $n>1$
are enumerated by those elements of the kernel of $C$ that {\em cannot\/}
be generated by any process of natural growth acting on 1-primitives of
lesser weight. This negative criterion appears even harder to implement
than the tensorial definition $U_1(X)=0$, which we were able to solve at
$n=9$, by explicit computation of the 98-dimensional
kernel of a $3214\times719$ matrix of integers.

Finally, we remark that we have {\em explicit\/} constructions
of the bigradings~(\ref{ficm},\ref{filad})
of the Connes-Moscovici and ladder subalgebras.
In the case of ${\cal H}_{\rm CM}$ we have merely a pair of 1-primitives:
$\delta_1=l_1$ and
$\widetilde\delta_2=N_{\delta_1}(\delta_1)-\frac12\delta_1^2$.
The only form of natural growth that we are allowed is by a single node:
this is the defining restriction.
Then we easily construct $\widetilde\delta_{k+1}
=N^k_{\delta_1}(\delta_1)-2^{-k}k!\delta_1^{k+1}$ as a nonproduct
$k$-primitive of weight $k+1$. This completes the filtration,
since any further term would make the number of weight-$n$ products
of filtered elements greater than the number of weight-$n$
products of the linear basis. Hence the construction of the
Connes-Moscovici bigrading is particularly simple.
In the case of ${\cal H}_{\rm ladder}$
the cocommutativity of the ladder restriction~(\ref{Dl}) of the
coproduct means that all $k$-primitives are products at $k>1$.
Here the problem of construction is more demanding, since it not
clear how to generate an infinite set of 1-primitives. Hence one
sees that detailed study of ladder diagrams, most notably by Bob
Delbourgo and colleagues~\cite{Del4,Del6,KD}, addresses a problem
more severe than that posed by the Connes-Moscovici prolegomenon
to noncommutative geometry: ladder diagrams are a nontrivial
infinite subset of perturbative quantum field theory; even after
subtractions of products they provide an infinite subset of
1-primitives, when their bigrading is analyzed. Fortunately, our
recent work in~\cite{BK2} provides the explicit
construction~(\ref{pl}) of the ladder filtration. The
reader may try to imagine what might be involved in giving an
explicit construction of the 1-primitives of the full Hopf algebra
of undecorated rooted trees. Then s/he should contemplate the true
challenge of quantum field theory, by recalling that -- in
physical reality -- every node of every rooted tree may be
decorated in an infinite number of ways. After half a century, few
physicists or mathematicians have even {\em begun\/} to grapple
with the true legacy of Dyson, Feynman, Schwinger and Tomonaga.

\section{Prospects}

In this paper, we were content to study the bigrading
of the Hopf algebra of undecorated~\cite{BK1,BK2} rooted trees,
by the number of nodes and a degree of primitivity analyzed by
iterations of the coproduct.
The extension of this bigrading to the decorated~\cite{Over,DR2}
case is the obvious next step, in our plan to decode the rich
structure of mature quantum field theory.
The present work makes it clear that the key feature will
be the nontriviality of
$C:=B_+\circ B_-\not=B_-\circ B_+={\rm id}$.
In the undecorated case, we have shown that the bifiltration of
the Hopf algebra is obtained by adjoining the single-node tree
to the bifiltration of the kernel of $C$.
The proof of this lies in the powerful generalization~(\ref{Vk})
of the concept of natural growth, which diagonalizes~(\ref{Ukm}).
First results for the commutator $[B_+,B_-]=C-{\rm id}$
of the decorated Hopf algebra of full quantum field theory
were recently given in~\cite{shuffle}. These increase our hopes
that it will not take another 50 years to complete the
characterization of the intricate interrelation of combinatorics {\em and\/}
analysis that makes quantum field theory possible. We firmly believe that
further elucidation of its structure has much to offer for wide areas
of both physics and mathematics.

\noindent{\bf Acknowledgements:}\quad
This study began during the
workshop {\sl Number Theory and Physics} at the ESI in November
1999, where we enjoyed discussions with Pierre Cartier, Werner
Nahm, Ivan Todorov and Jean-Bernard Zuber. Work with Alain Connes
at the IHES supports the present paper.
System management by Chris
Wigglesworth enabled accumulation of crucial data, which
Neil Sloane's {\tt superseeker} helped us to decode.

\section*{Appendix: asymptotic enumerations}

Here we consider inequalities inferred from Tables~3 and~4 and show that
they persist at large weights, thanks to the upper bound $c<3$
on the Otter constant~\cite{Otter}.

Asymptotically, the number of rooted trees is given by
\begin{equation}
r_n={c^n n^{-3/2}}(b+O(1/n))
\label{asy}
\end{equation}
with Otter constants that we evaluated in~\cite{BK1}:
\begin{verbatim}
   b = 0.43992401257102530404090339143454476479808540794011
         98576534935450226354004204764605379862197779782334...
   c = 2.95576528565199497471481752412319458837549230466359
         65953504724789059647331395749510866682836765813525...
\end{verbatim}
The asymptotic fraction of trees assigned to primitivity $k$
in the filtration of Table~4 is
\begin{equation}
f_k:=\lim_{n\to\infty} {P_{n,k}\over r_n}=
\left(1-{1\over c}\right)^{k-1}{1\over c}
\label{fk}
\end{equation}
while the asymptotic fraction of forests in Table~3 is
\begin{equation}
g_k:=\lim_{n\to\infty} {H_{n,k}\over r_{n+1}}=\frac{k f_k}{c}
=\left(1-{1\over c}\right)^{k-1}{k\over c^2}\,.
\label{gk}
\end{equation}
These follow by using~\cite{BK1} $|1-R(x)|^2=O(1-c x)$, near $x=1/c$.
Numerically,
\begin{eqnarray*}
g_1&=&
0.1144616788557279695\ldots\\
g_2&=&
0.1514735822429146084\ldots\\
g_3&=&
0.1503401379409753267\ldots\\
g_4&=&
0.1326357110750687024\ldots\\
g_5&=&
0.1097026887662558145\ldots\\
g_6&=&
0.0871054456752243543\ldots\\
g_7&=&
0.0672417311397409555\ldots\\
g_8&=&
0.0508484386279160206\ldots\\
g_9&=&
0.0378509630072558308\ldots
\end{eqnarray*}
with $k=2$ giving the largest fraction of forests at large $n$. This was
not apparent until $n=28$, where we found that
$H_{28,2}=20\,716\,895\,918$ exceeds $H_{28,3}=20\,710\,700\,277$.

The asymptotic results establish
inequalities~(\ref{keq2},\ref{kgt2}) at large $n$, where it is sufficient
that $c<3$.
Amusingly, this upper bound and the condition $R(1/c)=1$
produce a rather tight lower bound
\begin{equation}
c=\exp\left(\sum_{k>0}{R(c^{-k})\over k}\right)
>\exp\left(1+\sum_{k>1}{1\over(3^k-1)k}\right)>2.943
\label{lower}
\end{equation}
from the rather loose lower bound $R(x)\ge R_{\rm ladder}(x)=x/(1-x)$.
\raggedright

\newpage

\begin{center}{\bf Table~1}: Dimensions $\overline{H}_{n,k}$
of the bigrading the cocommutative
subalgebra, ${\cal H}_{\rm ladder}$\end{center}
$$\begin{array}{r|rrrrrrrrrrrrrrrrrrr}
&1&2&3&4&5&6&7&8&9&10&11&12&13&14&15&16&17&18&19\\\hline
1&1\\
2&1&1\\
3&1&1&1\\
4&1&2&1&1\\
5&1&2&2&1&1\\
6&1&3&3&2&1&1\\
7&1&3&4&3&2&1&1\\
8&1&4&5&5&3&2&1&1\\
9&1&4&7&6&5&3&2&1&1\\
10&1&5&8&9&7&5&3&2&1&1\\
11&1&5&10&11&10&7&5&3&2&1&1\\
12&1&6&12&15&13&11&7&5&3&2&1&1\\
13&1&6&14&18&18&14&11&7&5&3&2&1&1\\
14&1&7&16&23&23&20&15&11&7&5&3&2&1&1\\
15&1&7&19&27&30&26&21&15&11&7&5&3&2&1&1\\
16&1&8&21&34&37&35&28&22&15&11&7&5&3&2&1&1\\
17&1&8&24&39&47&44&38&29&22&15&11&7&5&3&2&1&1\\
18&1&9&27&47&57&58&49&40&30&22&15&11&7&5&3&2&1&1\\
19&1&9&30&54&70&71&65&52&41&30&22&15&11&7&5&3&2&1&1\\
\end{array}$$

\begin{center}{\bf Table~2}: Dimensions $\widetilde{H}_{n,k}$
of the bigrading the noncocommutative
subalgebra, ${\cal H}_{\rm CM}$\end{center}
$$\begin{array}{r|rrrrrrrrrrrrrrrrrrr}
&1&2&3&4&5&6&7&8&9&10&11&12&13&14&15&16&17&18&19\\\hline
1&1\\
2&1&1\\
3&&2&1\\
4&&1&3&1\\
5&&&2&4&1\\
6&&&1&4&5&1\\
7&&&&2&6&6&1\\
8&&&&1&4&9&7&1\\
9&&&&&2&7&12&8&1\\
10&&&&&1&4&11&16&9&1\\
11&&&&&&2&7&16&20&10&1\\
12&&&&&&1&4&12&23&25&11&1\\
13&&&&&&&2&7&18&31&30&12&1\\
14&&&&&&&1&4&12&27&41&36&13&1\\
15&&&&&&&&2&7&19&38&53&42&14&1\\
16&&&&&&&&1&4&12&29&53&67&49&15&1\\
17&&&&&&&&&2&7&19&42&71&83&56&16&1\\
18&&&&&&&&&1&4&12&30&60&94&102&64&17&1\\
19&&&&&&&&&&2&7&19&44&83&121&123&72&18&1\\
\end{array}$$

\newpage

\begin{center}{\bf Table~3}: Dimensions $H_{n,k}$
of the bigrading of the
Hopf algebra of rooted trees, ${\cal H}_{R}$\end{center}
$$\begin{array}{l|rrrrrrrrrrrrr}
&1&2&3&4&5&6&7&8&9&10&11&12&13\\\hline
1&    1\\
2&    1&    1\\
3&    1&    2&    1\\
4&    2&    3&    3&    1\\
5&    3&    6&    6&    4&    1\\
6&    8&   11&   13&   10&    5&   1\\
7&   16&   26&   27&   24&   15&   6&   1\\
8&   41&   58&   63&   55&   40&  21&   7&   1\\
9&   98&  142&  148&  132&  100&  62&  28&   8&   1\\
10&  250&  351&  363&  322&  251& 168&  91&  36&   9&  1\\
11&  631&  890&  912&  804&  635& 444& 266& 128&  45& 10&  1\\
12& 1646& 2282& 2330& 2051& 1625&1167& 742& 402& 174& 55& 11& 1\\
13& 4285& 5948& 6036& 5304& 4220&3072&2030&1184& 585&230& 66&12& 1
\end{array}$$

\begin{center}{\bf Table~4}: Filtration $P_{n,k}$
of the bigrading of ${\cal H}_{R}$\end{center}
$$\begin{array}{l|rrrrrrrrrrrr}
&1&2&3&4&5&6&7&8&9&10&11&12\\\hline
1&1\\
2&1\\
3&1&1\\
4&2&1&1\\
5&3&3&2&1\\
6&8&5&4&2&1\\
7&16&13&9&6&3&1\\
8&41&28&21&13&8&3&1\\
9&98&71&49&33&20&10&4&1\\
10&250&174&121&79&50&27&13&4&1\\
11&631&445&304&201&127&74&38&16&5&1\\
12&1646&1137&776&510&325&192&106&49&19&5&1\\
13&4285&2974&2012&1326&844&512&290&148&65&23&6&1\\
\end{array}$$

\end{document}